\documentclass[review]{elsarticle}
\usepackage[utf8]{inputenc}

\usepackage{listings}
\usepackage{color}
\usepackage[usenames,dvipsnames,svgnames,table]{xcolor}
\usepackage{bm}

\usepackage{algorithm}
\usepackage{array}
\usepackage{lineno,hyperref}
\usepackage{euscript}
\usepackage{amsthm, amssymb}
\usepackage{graphicx}
\usepackage{paralist}
\usepackage[noend,algo2e,lined,ruled,linesnumbered]{algorithm2e}
\usepackage{fancyvrb}
\usepackage{fvextra}
\usepackage{mathtools}
\usepackage{subfig}
\usepackage{multirow}
\usepackage{enumitem}
\usepackage{array,booktabs}
\usepackage{comment}
\usepackage{todonotes}
\usepackage{yhmath}

\usepackage{float}

\setlist[enumerate]{itemsep=0mm}

\usepackage{xifthen}
\usepackage{xparse}

\tikzstyle{place} = [
  circle,
  thick,
  draw=black,
  fill=white,
  minimum size=3mm,
  font=\fontsize{9}{144}\selectfont
]

\tikzstyle{transition} = [
  rectangle,
  thick,
  draw=black,
  fill=white,
  minimum width=2cm,
  minimum height=1cm
]

\tikzstyle{htransition} = [
  transition,
  fill=white,
  minimum width=8mm,
  minimum height=8mm,
]

\DeclareDocumentCommand{\Consts}{ o }{%
        \IfNoValueTF{#1}%
            {\textit{Const}_{\anet}}%
            {\textit{Const}_{#1}}%
}

\newcommand{\net}{DPN\xspace}

\newcommand{\anet}{\mathcal{N}}

\def\pre#1{\ensuremath{^\bullet{#1}}}
\def\post#1{\ensuremath{{#1}\kern-.05ex^\bullet}}

\def\posto#1{\ensuremath{{#1}\kern-.05ex\circ}}

\newcommand{\true}{\small{\texttt{true}}\xspace}
\newcommand{\false}{\small{\texttt{false}}\xspace}

\newcommand{\Vars}{V}
\newcommand{\VarsR}{\Vars^r}
\newcommand{\VarsW}{\Vars^w}

\newcommand{\A}{\mathcal{A}}

\newcommand{\nat}{\mathbb{N}}

\newcommand{\quasi}{\sqsubseteq}

\DeclareDocumentCommand{\typevar}{ o }{%
        \IfNoValueTF{#1}%
            {\var_\Domain}%
            {{\var_{#1}}_{\Domain}}%
}

\DeclareDocumentCommand{\var}{ o }{%
        \IfNoValueTF{#1}%
            {v}%
            {v_{#1}}%
}

\newcommand{\varInit}{\varState_I}
\newcommand{\reads}{\mathit{read}}
\newcommand{\writes}{\mathit{write}}
\newcommand{\guard}{\mathit{guard}}

\newcommand{\readvar}[1]{#1^r}

\newcommand{\varState}{\alpha}

\newcommand{\set}[1]{\{#1\}}                      
\newcommand{\tup}[1]{\langle #1\rangle}            

\newcommand{\cval}[1]{\mathtt{#1}}

\newcommand{\goto}[1]{\ensuremath[{#1}\rangle}
\newcommand{\arc}[1]{\xrightarrow{#1}}

\newcommand{\Domain}{\mathcal{D}}

\newcommand{\Reals}{\mathbb{R}}

\newcommand{\assignments}[1]{\ensuremath[\![#1]\!]}

\DeclareDocumentCommand{\CG}{ o }{%
        \IfNoValueTF{#1}%
            {CG_\anet}%
            {CG_{#1}}%
}
\DeclareDocumentCommand{\RG}{ o }{%
        \IfNoValueTF{#1}%
            {RG_\anet}%
            {RG_{#1}}%
}

\usepackage{color,soul}

\definecolor{incolor}{RGB}{210,220,230}
\definecolor{outcolor}{RGB}{235,215,215}

\tikzset{
table/.style={
  matrix of nodes,
  row sep=-\pgflinewidth,
  column sep=-\pgflinewidth,
  nodes={
    rectangle,
    draw=black,
    minimum width=.7cm,
    minimum height=5mm,
    align=center },
  text depth=0.25ex,
  text height=1ex,
  nodes in empty cells
  },
  dmn/.style={
    matrix of nodes,
    row sep=-\pgflinewidth,
    column sep=-\pgflinewidth,
    nodes={
      rectangle,
      draw=black,
      text width=12mm,
      minimum height=5mm,
      align=center },
    nodes in empty cells,
  },
  dmnhit/.style={
    rectangle,
    draw,
    minimum height=10.55mm,
    minimum width=7.1mm,
    xshift=1.3mm
  },
  dmnrulen/.style={
    matrix of nodes,
    row sep=-\pgflinewidth,
    column sep=-\pgflinewidth,
    nodes={
      rectangle,
      draw=black,
      text width=5mm,
      minimum height=5mm,
      align=center },
    nodes in empty cells,
  },
}

\usepackage{tikz}
\usetikzlibrary{arrows,shapes,shapes.multipart,snakes,automata,backgrounds,petri,positioning,shadows,matrix,decorations.pathmorphing, decorations.pathreplacing, decorations.markings, fit,positioning,calc,backgrounds,shapes.misc,arrows.meta,fit}

\graphicspath{{./Figures}}

\theoremstyle{definition}
\newtheorem{definition}{Definition}[section]
\newtheorem{example}{Example}[section]

\newtheorem{proposition}{Proposition}[section]

\begin{document}

\begin{frontmatter}
\title{Soundness Correction of Data Petri Nets}

\author[inst1]{Nikolai M. Suvorov\corref{cor1}} 
\ead{nmsuvorov@hse.ru}
\author[inst1]{Irina A. Lomazova} 
\ead{ilomazova@hse.ru}
\author[inst2]{Andrey Rivkin} 
\ead{ariv@dtu.dk}

\affiliation[inst1]{organization={HSE University},
            addressline={Myasnitskaya ul. 20}, 
            city={Moscow},
            postcode={101000}, 
            country={Russia}}
\affiliation[inst2]{organization={Technical University of Denmark},
            addressline={Richard Petersens Plads, Building 324}, 
            city={Kgs. Lyngby},
            postcode={2800}, 
            country={Denmark}}

\cortext[cor1]{Corresponding author}

\begin{abstract}
A process model is called sound if it always terminates properly and each model activity can occur in a process instance. 
Conducting soundness verification right after process design allows one to detect and eliminate design errors in a process to be implemented.
The process of eliminating such errors is called soundness repair.
In many repair scenarios, the resulting model should retain only the correct behavior of the source model, especially if a model is created manually.
In this paper, we consider this type of soundness repair applied to data-aware process models represented as data Petri nets (DPNs). Specifically, we investigate the capabilities to repair soundness of DPNs by restricting the transition guards and propose a new repair algorithm that follows this approach. A distinctive feature of the algorithm is the absence of a requirement for an input DPN to have a sound control flow.
The algorithm is implemented and results of the preliminary evaluation demonstrate its applicability to process models of moderate sizes.
\end{abstract}


\end{frontmatter}

\section{Introduction}
\label{sec:intro}

A business process comprises a set of coordinated activities performed within an organizational and technical environment to achieve a specific business goal~\cite{Weske2012}.
Analyzing business process models helps identify inconsistencies and vulnerabilities, providing valuable insights that can serve as a foundation for decision-making aimed at improving and optimizing these processes.

Real-world processes often rely on data that is manipulated by process activities and referenced at various decision points. 
Modeling support for data access and manipulation is provided, for instance,  by the BPMN 2.0 standard\footnote{\url{https://www.omg.org/spec/BPMN/2.0/}} or various data-aware process formalisms and frameworks (e.g., \cite{ChiaoKR13,HaarmannMW21,GhilardiGMR23,GhilardiGMR21,PolyvyanyyWOB19,ReijersVPGFCG17,MeyerPFW13,CarrasquelLR20,AlmanMRRW24,LiDV17,BourhisH0S20,Leoni_2013,Hildebrandt0S23}). 
However, the resulting process models are not inherently guaranteed to be ``flawless''. 
At the design stage, numerous errors can arise at both the control-flow level (e.g., deadlocks caused by poorly implemented mutexes) and the data manipulation level (e.g., inconsistent data updates or incorrect checks at decision points).
Such errors can be addressed by either verifying process models against a set of (temporal) properties using Model Checking~\cite{mc-book} or 
checking more holistic, business process-specific properties such as \emph{soundness}~\cite{Aalst_11}.

In this work, we focus on a specific formalism for data-aware processes called Data Petri Nets (DPNs)~\cite{Leoni_2013}. Data Petri Nets extend standard place/transition nets with data manipulation capabilities, enabling transitions to perform checks and updates on a fixed set of (typed) variables. Thus, each state in a DPN is represented as a pair consisting of the standard net marking and variable valuations (i.e., values assigned to all the variables). As shown in~\cite{LeoniFM21}, DPNs can be used as a formal counterpart of a fragment of BPMN enriched with decision tables. 
Properties like soundness have been also studied in the context of Data Petri Nets. A DPN is called \emph{data-aware sound} if it always terminates with respect to some variable valuation and each activity can occur along a net execution~\cite{deLeoni18}. Compared to classical soundness, where only the control flow is investigated, data-aware soundness captures the interplay of control and data flows simultaneously.

Several algorithms~\cite{Felli19,Felli2021,Felli2022,Suvorov2024} have been proposed for verifying the data-aware soundness of Data Petri Nets. However, an important question arises: what actions can be taken when a model is identified as unsound? 
Once the sources of unsoundness have been identified, one may want to attempt to eliminate them, ensuring that the underlying process model becomes sound. 
Model repair is currently a predominantly manual effort, as the problem of repairing unsound data-aware models has not yet been widely investigated in the research community.
At the time of writing, two soundness repair algorithms for DPNs have been proposed, namely those described in~\cite{Zavatteri_2024} and~\cite{Felli_Repair}.
However, both algorithms are primarily designed to be used in process mining scenarios, where DPN models are discovered in a two-step process. First, the control flow of the target model is discovered using algorithms that ensure the soundness of the resulting "backbone" Petri net (e.g., Inductive Miner~\cite{InductiveMiner}, Evolutionary Tree Miner~\cite{ETM}, Structured Miner~\cite{STM}). Second, data guards are discovered and added to the corresponding transitions. At this stage, any unsoundness in the model can only stem from the data guards.

In this paper, we consider a more general case of DPN soundness repair, where the model may either be built manually or discovered using a process discovery algorithm that does not guarantee any form of soundness.
We ensure that the repaired model contains only the behavior that already leads to proper termination in the original model, thereby guaranteeing that no new behavior is introduced during the repair process. 
This guarantee aligns with the `natural' intuition behind manually designed models: a domain expert often creates an accurate representation of the correct (or desired) process executions but may overlook subtle issues such as deadlocks, livelocks, or unbounded resource growth. In such cases, adding new behavior is typically not desired.

\begin{figure}[htp]
		\centering
        \includegraphics[width=0.5\textwidth]{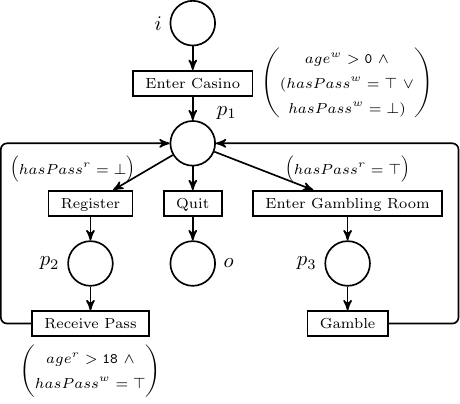}
		\caption{DPN $\anet$ representing a visit to a casino with a deadlock. $M_I = [i]$ and $M_F = [o]$. Variable \textit{age} is of a real type and initialized to $\cval{0}$. Variable \textit{hasPass} is of a boolean type and initialized to \false.}
		\label{fig:casino}
	\end{figure}

A straightforward approach to repairing the system in this context is to restrict its transition guards, thereby preventing the undesired behavior.
Figure~\ref{fig:casino} shows an example of a process model for which a meaningful repair involves restricting transition guards. 
The model depicts a visit to a casino and runs into a deadlock when an individual under 18 years old registers for a casino pass (which also makes the model unsound). By adopting an approach that restricts transition guards, we can prohibit minors from applying for a pass, thereby making the model sound. Conversely, if we take an approach that relaxes transition guards, we would need to relax the guard of the Receive Pass transition by replacing $\readvar{age} > \cval{18}$ with $\readvar{age} > \cval{0}$, which would allow minors to gamble in the casino -- an outcome that is clearly undesirable.

In this work, we investigate an approach that repairs unsound DPNs by restricting some of the transition guards. As opposed to~\cite{Felli_Repair}, the approach does not require a sound control flow and often requires significantly fewer abstract state space constructions, owing to a different soundness verification procedure~\cite{Suvorov2024}.
As input, we consider DPNs with transition guards expressed as boolean combinations of atomic variable-operator-variable and variable-operator-constant formulas, where all components are real-typed. This formulation is critical to ensuring the decidability of the data-aware soundness verification task~\cite{Suvorov2024}.

The main research contributions of this work are as follows:
\begin{compactenum}
    \item The definition of the applicability boundaries for soundness repair algorithms (for DPNs) based on restricting transition guards.
    \item A soundness repair algorithm for DPNs with both sound and unsound control flows, for which we formally show the termination result. 
    \item A fully-fledged research prototype implementing the proposed algorithm, accompanied by a preliminary experimental evaluation on synthetically generated models. This evaluation highlights the potential applicability of our algorithm in real-world scenarios and includes an execution time comparison with the algorithm presented in~\cite{Felli_Repair}. 
\end{compactenum}

The remainder of the paper is organized as follows. Section~\ref{sec:dpns} provides the syntax and semantics of DPNs. 
Section~\ref{sec:applicability}  examines the applicability boundaries of soundness repairs for DPNs based on restricting transition guards.
Section~\ref{sec:repair_algorithm} introduces the soundness repair algorithm. 
Section~\ref{sec:implementation} shows the prototype implementation of the algorithm together with its preliminary performance evaluation. Section~\ref{sec:related_work} discusses the related work and existing soundness repair algorithms. Section~\ref{sec:conclusion} concludes the paper.

\section{Data Petri Nets}
    \label{sec:dpns}

    In this chapter, we describe the syntax and semantics of Data Petri nets (DPNs) and introduce a notion of data-aware soundness. 

    As already mentioned in Section~\ref{sec:intro}, DPNs is an extension of Petri nets with data variables. DPN transitions represent activities and are associated with guards that define input and output conditions over the data variables.
    
	We define a language of \emph{arithmetic constraints} capable of representing such input/output conditions imposed by process activities. The language described here is also used further to define a language of state constraints.

\begin{definition}[Arithmetic constraint]
  \label{def:constraint-syntax}
An \emph{arithmetic constraint} $\varphi$ over a set $X$ of  variables is an expression of the form:
\[\varphi :=  \top \mid x\odot y \mid x\odot c \mid\lnot\varphi\mid\varphi_1\land\varphi_2,\]
where:
	\begin{inparaenum}[(i)]
		\item $\top$ is the logical ``true'';
            \item $x,y \in X$;
		\item $c\in \Reals$;
		\item ${\odot}\in\set{<,=,>}$.
	\end{inparaenum}
\end{definition}
In the following, we make use of the following standard equivalences:
\begin{inparaenum}[(i)]
\item $\neg \top = \bot$ (logical ``false'');
\item $\varphi_1\lor\varphi_2 = \lnot(\lnot \varphi_1\land\lnot\varphi_2)$;
\item $x\leq y = \lnot(x>y)$; 
\item $x\geq y = \lnot (x<y)$; and 
\item $x\neq y = \lnot (x=y)$.
\end{inparaenum}
We denote by $\Phi(X)$ the language of arithmetic constraints over variables from $X$. 
For example, for $X = \set{y,z}$, all the following formulas are in $\Phi(X)$: $y < z$, $z \neq 3$, $(y \ge 3) \land (z > y)$, $(z > 1) \lor ((z \le 2) \land (y = 1))$.

We now formalize the interpretation of arithmetic constraints.
\begin{definition}[Satisfaction of an arithmetic constraint]
\label{def:constraint-semantics}
Given a set $X$ of variables, an arithmetic constraint $\varphi \in \Phi(X)$ is \emph{satisfied} by an assignment $\theta: X \rightarrow \Reals$, written $\theta \models \varphi$, if the following conditions hold:
	\begin{compactitem}
		\item $\theta \models x \odot c$ iff $\theta(x)$ is defined, and $\theta(x)\odot c$ is true; 
            \item $\theta \models x \odot y$ iff both $\theta(x)$ and $\theta(y)$ are defined, and $\theta(x)\odot \theta(y)$ is true;
		\item $\theta\models\lnot\varphi$ iff $\theta\not\models\varphi$;
		\item $\theta\models\varphi_1\land\varphi_2$ iff $\theta\models\varphi_1$ and
		$\theta\models\varphi_2$.
	\end{compactitem}
\end{definition}
We denote by $\assignments{\varphi}$ the set of all possible assignments that satisfy $\varphi\in \Phi(X)$.
Formally, $\assignments{\varphi}\doteq\{\theta \mid \theta\models\varphi\}$.
We say that two formulas  $\varphi_1, \varphi_2 \in \Phi(X)$ are \emph{logically equivalent} (denoted $\phi_1 \sim \phi_2$) iff 
$\assignments{\varphi_1}=\assignments{\varphi_2}$.

\smallskip
\noindent\textbf{DPN syntax.}
Data Petri nets (DPNs)~\cite{deLeoni18,Mannhardt2016} extend traditional place-transition nets with the possibility of manipulating scalar net variables from a given set $V$ that are also used to constrain the net evolution via
\emph{guards} assigned to net transitions. 
For each variable $v\in V$, we introduce additional symbols $v^r$ and  $v^w$ respectively used to denote input and output values of $v$.
Without loss of generality, we introduce two sets $\VarsR \doteq \set{ \var^r \mid \var\in\Vars }$ and $\VarsW \doteq \set{ \var^w \mid \var\in\Vars }$ storing the above symbols. 
Like that, each guard is an arithmetic constraint from $\Phi(\VarsR\cup \VarsW)$.

\begin{definition}[Data Petri net]
\label{def:dpn}
A \emph{data Petri net} (DPN) is a tuple
$\anet = \tup{P, T, F, V, \guard}$, where:
\begin{compactenum}[\it (i)]
			\item $P$ and $T$ are disjoint sets of places and transitions, respectively;
			\item $F:(P \times T) \cup (T \times P)\to\nat$ is a flow relation;
			\item $\Vars$ is a finite set of variables;
			\item $\guard: T \rightarrow \Phi(\VarsR\cup\VarsW)$ is the \emph{guard} assignment function labeling transitions with arithmetic constraints.
\end{compactenum}
\end{definition}

Given a DPN $\anet = \tup{P, T, F, V, \guard}$, 
we will write $P_{\anet}$, $T_{\anet}$, etc. to denote $\anet$'s components; we omit the subscript if the referenced net is clear from the context. 
Given a place or transition $x\in (P_{\anet} \cup T_{\anet})$ of $\anet$, the \emph{preset} $\pre{x}$ and the
\emph{postset} $\post{x}$ are given by
$\pre{x}=\set{y\mid (y,x)\in F}$ and
$\post{x}:=\set{y\mid (x,y)\in F}$.
Given $t\in T$, we also define $\reads(t)$ and $\writes(t)$ to denote, respectively, all the variables from $\VarsR$ and $\VarsW$ that occur in $\guard(t)$.

\smallskip
\noindent\textbf{DPN execution semantics.} 
A \emph{state} of a DPN $\anet$ is a pair $(M,\alpha)$, where
\begin{compactenum}[\it(i)]
	\item $M:P_\anet\to\nat$ is a total \emph{marking} function, assigning a number $M(p)$ of \emph{tokens}
		to every place $p\in P_N$ and
	\item $\alpha:V_{\anet}\to\Reals$ is a total \emph{variable valuation} function assigning a value to every variable in $V_{\anet}$.
\end{compactenum}
We use $\A$ to denote the set of all possible variable valuations.
When variable valuations are not important in a given context, we shall talk about markings instead of states. 
Given two markings $M'$ and $M''$ of a \net $\anet$, we write $M'' \succeq M'$ iff for all $p \in P_\anet$ we have $M''(p) \geq M'(p)$, and we write $M'' \succ M'$ iff $M'' \succeq M'$ and there exists $p \in P_\anet$ s.t.\ $M''(p) > M'(p)$.
We use $\mathcal{M}_\anet$ to denote all markings of $\anet$.

A DPN moves between states by firing (enabled) transitions. 
After a transition fires, a new state
is reached, with a new corresponding marking and valuation.
    
	\begin{definition}[Transition firing]
		\label{def:transfiringnew}
		Given a DPN $\anet$ and some state $(M,\varState)$, we say that transition $t\in T$ may \emph{fire} at $(M,\varState)$ yielding a new state $(M',\varState')$ iff:
		\begin{compactitem}
			\item $M(p) \ge F(p,t)$ and $M'(p) = M(p) - F(p,t) + F(t,p)$, for all $p\in P$;
			\item $\beta\models\guard(t)$, where  $\beta:\VarsR\cup\VarsW\to \Reals$ and, for every $v\in V$, it holds that $\beta(v^r)=\alpha(v)$ and $\beta(v^w)=\alpha'(v)$;
                \item $\alpha(v)=\alpha'(v)$, for every $v\in V$ such that $v^w\not\in\writes(t)$.
		\end{compactitem}
        We denote transition firing as $(M,\alpha)\goto{t}(M',\alpha')$.
	\end{definition}
	
	The above definition can be easily extended to finite
 sequences of transition firings $\sigma = t^1\cdots t^n$, called \emph{traces}. 
 A trace, in turn, induces a (net) \emph{run} denoted as 
	$(M^0,\varState^0) \goto{t^1} \ldots \goto{t^n} (M^n,\varState^n)$ (or, equivalently, as $(M^0,\varState^0)\goto{\sigma}(M^n,\varState^n)$).
Given two states $(M,\varState)$ and $(M',\varState')$, we will also write $(M,\varState) \goto{*} (M',\varState')$ for cases in which $(M,\varState)=(M',\varState')$ or when there exists a trace $\sigma$ s.t. $(M,\varState) \goto{\sigma} (M',\varState')$.

 \begin{definition}[Reachability set, reachability graph]
     Given a DPN $\anet$ with an initial state $(M_I, \alpha_I)$.
     The \emph{reachability set} of $\anet$, denoted as $Reach_{\anet}$, is the smallest set of states that is inductively defined as follows:
     \begin{compactitem}
		\item $(M_I, \alpha_I)\in Reach_{\anet}$;
		\item if $(M, \alpha) \goto{t} (M', \alpha')$ for $t \in T$ and $(M, \alpha)\in Reach_{\anet}$, then $(M', \alpha') \in Reach_{\anet}$.
	\end{compactitem}
 The \emph{reachability graph} of $\anet$, denoted as $\RG$, is a graph $\tup{V,E}$, where:
 \begin{compactitem}
			\item $V = Reach_{\anet}$ is the set of reachable states of $\anet$; 
			\item $E\subseteq V\times T\times V$ is the set of edges such that $(v,t,v')\in E$ iff $v\goto{t} v'$, for some $t\in T$.
		\end{compactitem}
  \label{def:reachability}
 \end{definition}

    In the following, we will be interested in the boundedness property of DPNs.
    We say that a DPN $\anet$ is \emph{bounded} if there exists a bound $k \in \nat$ such that $M(p) \le k$, for all $p\in P$ and $(M,\alpha) \in Reach_{\anet}$. 

\begin{figure}[!ht]
		\centering
        \includegraphics[width=0.8\textwidth]{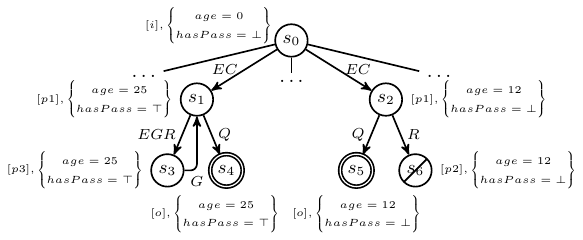}
		\caption{A fragment of the reachability graph for $\anet$ from Figure~\ref{fig:casino}. Arcs are labeled with the initials of the transition names. Square brackets denote markings. Curly brackets denote variable valuations. Double circles denote final nodes. Forbidden signs denote deadlocks.}
		\label{fig:rg}
	\end{figure}

    \begin{example}
        Consider DPN $\anet$ from Figure~\ref{fig:casino}.
        Initially, $age=\cval{0}$, and $hasPass=\bot$. At $(M_I,\alpha_I)$, only \textit{Enter Casino} may fire updating the values of $age$ and $hasPass$, so that $age$ becomes greater than $\cval{0}$ and $hasPass$ becomes either $\top$ or $\bot$. After that, multiple transitions may fire. \textit{Quit} may fire given any variable values leading to the final marking. \textit{Register} may fire only if $hasPass=\bot$. \textit{Enter Gambling Room} may fire only if $hasPass=\top$. If \textit{Register} fires, then only \textit{Receive Pass} may fire requiring $age$ to be greater than $\cval{18}$. This transition firing leads to the above decision point and updates $hasPass$ assigning it $\top$. If \textit{Enter Gambling Room} fires, then only \textit{Gamble} may fire also leading to the above decision point. 
        A fragment of the reachability graph for the DPN from Figure~\ref{fig:casino} is depicted in Figure~\ref{fig:rg}.
    
    \end{example}
    	
	\subsection{Data-aware soundness}

    Data-aware soundness is, perhaps, one of the key correctness criteria for DPNs that has been studied in-depth since the introduction of the formalism in \cite{deLeoni14_Conformance}. This criterion is similar to soundness for WF-nets~\cite{Aalst_11}, but instead of quantifying only over the reachable markings of the net, it also takes into account the states of the net variables. Below, we provide the definition of the data-aware soundness.
	
	\begin{definition}[Data-aware soundness~\cite{deLeoni18}]
	\label{def:soundness}	
		Let $\anet$ be a DPN with initial state $(M_I, \alpha_I)$ and final marking $M_F$. We say that $\anet$ 
            is \emph{data-aware sound} iff the following conditions hold:
		
		\begin{compactitem}
                \item[\textbf{C1}] for each $(M, \varState)\in Reach_{\anet}$, there exists $\varState'$ s.t. $(M,\varState) \goto{*} (M_F,\varState')$.
			\item[\textbf{C2}] for each $(M, \varState)\in Reach_{\anet}$, if $M\succeq M_F$ then $M=M_F$.
			\item[\textbf{C3}] for each $t\in T$, there exist $(M_1, \varState_1)$ and $(M_2,\varState_2$) such that $(M_1,\varState_1)\in Reach_{\anet}$ and $(M_1,\varState_1)\goto{t} (M_2,\varState_2)$.
		\end{compactitem}
	\end{definition}
	
	The first condition states the final state can be always reached. 
 The second condition captures that when the final state is reached, there should be no extra tokens in the net but those assigned by $M_F$. The last condition requires the absence of dead transitions.

\begin{example}
   Consider the DPN $\anet$ from  Figure~\ref{fig:casino}.
   This net is not data-aware sound as condition \textbf{C1} from Definition~\ref{def:soundness}  does not hold. 
   Indeed, if $age$ is assigned to a value not greater than $\cval{18}$ and \textit{hasPass} is assigned to $\bot$ after firing \textit{Enter Casino}, then firing \textit{Register} at $M=[p_1]$ leads to a situation when neither of transitions may fire. 
   A sample run that leads to this deadlock (and, thus, violates \textbf{C1}) is illustrated in the fragment of the reachability graph presented in Figure~\ref{fig:rg}. Specifically, this run is $s_0\goto{EC}s_2\goto{Q}s_6$, where \textit{Enter Casino} assigns $\cval{12}$ to $age$, and $\bot$ to $hasPass$.
\end{example}

In the above example, we saw one of the cases when a DPN can be unsound. 
Naturally, one may wonder whether soundness can be recovered by, for example, manipulating the data flow of the net. In the following section, we investigate the capabilities of soundness repair approaches based on restricting transition constraints.

\section{Limitations of the Transition Guards Restriction Approach}
  \label{sec:applicability}

  The approach that we investigate in this paper is based on restricting transition guards. Although this approach allows to save only correct behaviors of the input DPN,
  there are some cases in which it is either not applicable or can forbid some correct behaviours of the input DPN.

  \begin{figure}[!htb]
		\centering
        \includegraphics[width=.7\textwidth]{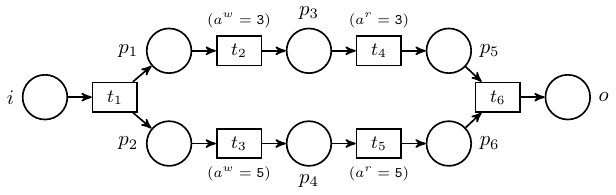}
		\caption{DPN $\anet$ with an initial state $M_I = [i]$ and $\alpha_I(a) = 0$, and a final state $M_F = [o]$. The net has a sound control flow that cannot be repaired by restricting transition guards.}
		\label{fig:counterexample}
	\end{figure}
  Consider DPN $\anet$ from Figure~\ref{fig:counterexample}. 
  This DPN has a sound control flow, i.e., only the part of the net without guards (this corresponds to the classical soundness notion from~\cite{Aalst_11}). 
  However, it is impossible to restrict the transition guards of this net to make it data-aware sound as per Definition~\ref{def:soundness}. 
  Note that transitions $t_2$ and $t_3$ cannot fire sequentially in any of the net executions leading to $M_F=[o]$: that kind of firing would prohibit the firing of either $t_4$ or $t_5$. This sequential execution cannot be forbidden only by restricting transition gaurds. The reason is that there is no input condition that can be additionally put on $t_2$ (or $t_3$, respectively) that will hold only after firing of $t_5$ (or $t_4$, respectively).
  This case can be generalized to DPNs with a sound control flow exhibiting concurrent behaviours obtained by splitting model executions into branches using (AND-split) and then joining them together (AND-join), and where at least two of such branches first update and then test  for equality (in different transitions) the same variables using different values.  
  For such nets, it is not always possible to properly order the transition firing by solely restricting transition guards. 

  \begin{figure}[!htb]
		\centering
        \includegraphics[width=.45\textwidth]{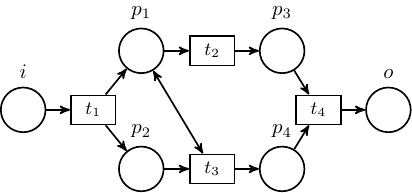}
		\caption{Bounded DPN $\anet$ with an unsound control flow that cannot be repaired by restricting transition guard. $M_I = [i]$ and $M_F = [o]$. For each transition, the guard is $\true$. }
		\label{fig:bounded_unrepairable}
	\end{figure}

  \begin{figure}[!htb]
		\centering
        \includegraphics[width=.3\textwidth]{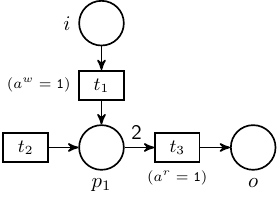}
		\caption{Unbounded DPN $\anet$ that cannot be repaired by restricting transition guards. Here, $M_I = [i]$ and $\alpha_I(a) = 0$, and $M_F = [o]$.}
		\label{fig:unbounded_unrepairable}
	\end{figure}
  Consider a bounded DPN with an unsound control flow in Figure~\ref{fig:bounded_unrepairable}. Here, the net does not have any input/output conditions on transitions. In this example, we can only switch guards of transitions to $\false$, but this cannot repair the net, since for its proper termination, each DPN transition must fire. Thus, without adding new writes to the transitions, this net cannot be repaired. This can also be the case for unbounded DPNs, for instance, for the DPN from Figure~\ref{fig:unbounded_unrepairable}. 
  To make the latter net sound, $t_2$ must be allowed to fire only once, but this cannot be done without adding a new output condition on $t_2$.

\begin{figure}[!htb]
		\centering
        \includegraphics[width=.5\textwidth]{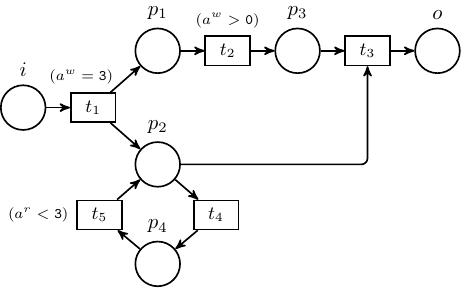}
		\caption{DPN $\anet$ with a sound control flow for which it is impossible to save all correct behavior when repairing by restricting transition guards. $M_I = [i]$ and $M_F = [o]$, $\alpha_I(a) = 0$.}
		\label{fig:behavior_loss}
	\end{figure}
 
    We have also found out that even for some DPNs with sound control flows it is impossible to preserve all correct executions of the source net in its repaired version if a repair is done by restricting transition guards. 
    The example is shown in~    Figure~\ref{fig:behavior_loss}. 
    This net reaches a deadlock if $t_2$ assigns $a$ a value greater than $3$ and then $t_4$ fires.
    Note that restricting guards of $t_1$ and $t_3$ would not anyhow help to make the model sound. Guard restriction of any transition from $\set{t_2,t_4,t_5}$ that makes the model sound forbids some of the correct executions. As an example, let us restrict the guard of $t_4$. To avoid the deadlock at $p_4$, the guard should be $a^r < c'$, where $c' \le 3$. This forbids a correct execution when $t_4$ fires with $a = 3$ and then $t_2$ updates $a$, so that it becomes less than 3. It is easy to see that an attempt to simultaneously restrict multiple transition guards of this net will also lead to a loss of the correct behavior. This case can be generalized to DPNs with sound control flows whose execution at some point splits into several concurrent threads, and at least one of these threads updates at least one variable $x$ while another thread has a transition $t$ such that $x\in\reads(t)$ (that is, $x$ is tested for some values in $guard(t)$).

  The mentioned above examples show the limitations of the approach that we investigate in this paper. The consequence of these limitations is the fact that any soundness repair algorithm that follows this approach cannot be complete even for the DPNs with a sound control flow. The same is true for the algorithm that we present in the following section. Note that these limitations do not make the investigated approach inapplicable: there exists a big portion of DPNs that still can be repaired following this approach. For DPNs with a sound control flow, the repair can always be done if a model does not have parallel execution threads. 
  In the case of parallel threads, we suppose that the investigated repair approach is not applicable only for DPNs, where at least two threads update the same variable and at least one thread checks its value. One could implement a graph-traversing algorithm that checks this (or stronger) condition before applying a soundness repair algorithm that follows the guard restriction approach to be sure that the algorithm would return a repaired net.

  It is also important to mention that the limitations demonstrated in this section's examples also appear in the algorithm based on restricting transition guards presented in~\cite{Felli_Repair}. Although the authors state that each DPN with a sound control flow that has at least one correct execution can be successfully repaired by restricting transition guards with the preservation of all correct behaviors of the source net, we have shown that this cannot always be true. 

\begin{figure}[!htb]
		\centering
        \includegraphics[width=.8\textwidth]{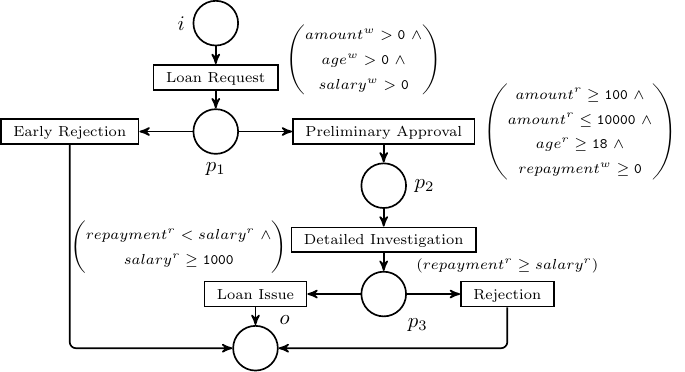}
		\caption{A DPN representing the process of getting a loan from a bank. Proper repair of this model requires specific domain knowledge. $M_I = [i]$ and $M_F = [o]$. All variables are of a real type and initialized to 0.}
		\label{fig:bpmn-new}
	\end{figure}

  Lastly, it is important to highlight the fact that in some situations it may not be desired to restrict the transition guards as it may slightly modify the business logic. Figure~\ref{fig:bpmn-new} shows a DPN that models the process of getting a loan from a bank. By following the restricting guards approach, we should restrict the guard of \textit{Preliminary Approval} so that it cannot fire if $repayment < salary \land salary < 1000$. However, in real process execution, it may be expected to go through \textit{Preliminary Approval} and \textit{Detailed Investigation} in this case and then to receive a rejection. If it is the expected behavior for a process, then the straightforward approach is to relax the guard of \textit{Rejection}. Proper repair of such models usually requires specific domain knowledge. In these situations, a modeler can use multiple repair approaches and select the result that best fits the domain.

\section{Soundness Repair Algorithm}
 \label{sec:repair_algorithm}

    In this section, we propose an algorithm for repairing data-aware soundness of a DPN. 
    In the nutshell, the algorithm iteratively refines a DPN, constructs a coverability graph for the refined DPN, and forbids unfeasible runs that lead to deadlocks, livelocks, or unboundedness.
    When all unintended behavior is removed by the iterative algorithm, dead transitions and isolated places are deleted, refined transitions are merged back, and the repaired model is returned to the user.

To introduce the algorithm, we first need to define a Labeled Transition System (LTS), a Coverability Graph, a Refined DPN, and a $\tau$-DPN.

    \subsection{Labeled Transition Systems and Coverability Graphs for Data Petri Nets}
    
    The repair algorithm we introduce in this section requires two additional structures that 
    we formally define below.

    First, we define a \emph{labeled transition system} (LTS for short) \emph{induced} by a DPN.
    Such a transition system can be seen as a generalization of a reachability graph (as per Definition~\ref{def:reachability}): 
    instead of representing a single DPN state, each node in a TS carries a set of states that have the same marking but different variable valuations. 
    Our definition of an LTS is equivalent to the definition of a constraint graph from~\cite{Felli2022} and~\cite{Felli_Repair}. We decided to propose a separate notion to distinguish it from a notion of a constraint graph defined in works~\cite{Felli19} and~\cite{Felli2021}, where a constraint graph is actually a labeled transition system that is induced not by DPN $\anet$ but by $\tau$-DPN $\anet_\tau$ (introduced later in Section~\ref{sec:process-repairs}).

    Recall that each DPN transition $t$ defines a non-deterministic transformation of the input variable valuation $\alpha$ into the output one.
    All such transformations can be characterized by the set $\rho(t,\alpha)=\bigl\{\beta\in\set{\VarsR\cup\VarsW\to \Reals}~\mid~ \beta\models\guard(t), \beta(v^r)=\alpha(v)\text{ for all }v\in V\bigr\}$.
    We also assume that $\rho$ can be extended to a set of variable valuations $A\subseteq\A$ as follows: 
    $\rho(t,A)=\set{\beta\in\rho(t,\alpha)\mid \alpha\in A }$.

    Now we can define an LTS induced by a DPN $\anet$.

    \begin{definition}[Labeled Transition System Induced by a \net]
		\label{def:LTS_new}
		Let $\anet$ 
  be a \net. A \emph{labeled transition system} $LTS_\anet$ \emph{induced by} $\anet$ is a tuple $\tup{S, E, s_0}$, where:
		\begin{compactitem}
			\item $S \subseteq \mathcal{M}_\anet \times 2^{\mathcal{A}_\anet}$ is a set of nodes;	
			\item $E \subseteq S \times T \times S$ is a set of arcs labeled with transitions s.t. a triple $\bigl((M,A),\,t,\,(M',A')\bigr)\in E$ iff:\footnote{We will denote node-edge-node triples as $(M,A)\arc{t}(M',A')$.}
			\begin{compactitem}
				\item $M(p) \ge F(p,t)$ and $M'(p) = M(p) - F(p,t) + F(t,p)$, for each $p\in P$;
                \item $A' = \rho(t,A)$ and $A'\neq\emptyset$.
			\end{compactitem}
			\item $s_0 = (M_I, A_I) \in S$ is the initial node with $A_I = \set{\alpha_I}$.
		\end{compactitem}
	\end{definition}

    Some of the LTS states may contain infinitely many variable valuations. 
    To account for this problem, we symbolically abstract each 
    such set from $(M,A)\in S$ using an arithmetic constraint $\phi$ from $\Phi(V)$.
    Like that, each $\phi\in \Phi(V)$ represents conditions imposed on values of variables from $V$, and every state in $LTS_\anet$ can be replaced with $(M,\phi)$.
    Language $\Phi(V)$ is sufficient to represent all possible variable valuations for the DPN setting considered in this paper but may not be sufficient for other DPN settings. 
    Results reported in~\cite{Suvorov2024} provide more detail on the aforementioned expressiveness problem.

 \begin{figure}[!ht]
    \centering    					
    \includegraphics[width=.8\textwidth]{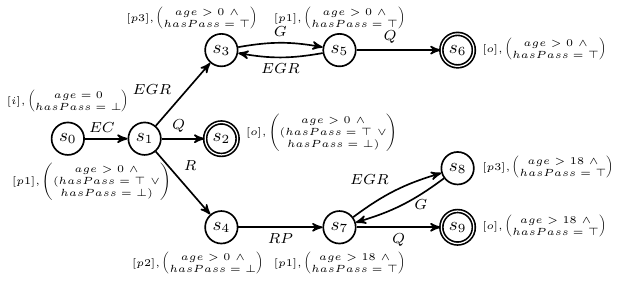}
    \caption{$LTS_\anet$ constructed for DPN $\anet$ from Figure~\ref{fig:casino}. Double circles denote final nodes.}
    \label{fig:casino_dpn_lts}
    \end{figure}
    
\begin{example}
    \label{ex:lts}
    Figure~\ref{fig:casino_dpn_lts} illustrates an $LTS_{\anet}$ constructed for DPN $\anet$ from Figure~\ref{fig:casino}. Consider sample DPN state $([p1], \set{age = 25, hasPass = \top})$. This state is abstracted by nodes $\set{s_1,s_5,s_7}$. A union of the incoming arcs to these nodes denotes the set of transitions whose firings may lead to this state. An intersection of the outgoing arcs from these nodes denotes the superset of transitions that may fire from this state. The latter is the reason why the final markings are reachable from any node of $LTS_{\anet}$, although $\anet$ is not data-aware sound.
\end{example}

    An LTS is a fairly memory-intensive model. 
    There exists some promising research dedicated to reducing the space needed for model verification, such as~\cite{GROEFSEMA2020103181}, but it mainly focuses on way simpler rules that only require few execution traces. Soundness, on the contrary, is a universal property requiring computational structures capturing all the possible behaviors of the system. Making the state space even smaller could be possible, but requires better study of heuristics that could help with it.
   
    We now move to defining a coverability graph of an LTS induced by a DPN. 
    First, we need to introduce the notion of coverability and define the quasi-ordering relation on LTS states.
	
    \begin{definition}[Coverability]
    \label{def:covers}
    Let $LTS_\anet$ be an LTS induced by a DPN $\anet$.
    Let $(M,A),(M',A')\in S_{LTS_\anet}$. 
    We say that $(M',A')$ \emph{covers} (resp., \emph{strictly covers}) $(M,A)$, denoted as $(M,A)~\quasi~ (M',A')$ (resp.,  $(M,A)~\sqsubset~ (M',A')$), iff $A = A'$ and $M \preceq M'$ (resp., $M \prec M'$).
    \end{definition}
    It is easy to see that $\quasi$ is a quasi-ordering relation (that is, it is reflexive and transitive). Now we can define a coverability graph.

    \begin{definition}(Coverability Graph of an LTS)
    \label{def:c-graph}
    Let $\anet$ be a \net and $LTS_\anet = \tup{S, E, s_0}$ be an LTS of $\anet$. 
    A \emph{coverability graph of  $LTS_\anet$} is $CG_{LTS_\anet} = \tup{S_{CG}, E_{CG}, s_{0}}$ such that:
    \begin{compactitem}
        \item $S_{CG} \subseteq S$ is the set of non-isolated nodes, where each node is classified as either \emph{dead} or \emph{live} as follows: 
        \begin{compactitem}
            \item $s'\in S_{CG}$ is  \emph{dead} if $s'$ does not have successors in $LTS_\anet$, or
                there exists a node $s \in S_{CG}$ along the path from $s_0$ to $s'$, s.t. $s \sqsubset s'$ (i.e., $s'$ strictly covers $s$);
            \item $s'\in S_{CG}$ is a live node, otherwise. 
        \end{compactitem}
        \item $E_{CG} \subseteq E$ is the set of arcs labeled with transitions $t\in T_\anet$, where $(s,t,s') \in E_{CG}$ iff the following holds:
            \begin{compactitem}
                \item $(s,t,s') \in E$;
			\item $s$ is a live node or the initial node.
			\end{compactitem}
        \item $s_0$ is the initial node.
    \end{compactitem}
    \label{def:ct}
    \end{definition}

    \smallskip

    Research \cite{Suvorov2024} proved that for a DPN $\anet$ as per Definition~\ref{def:dpn} $LTS_\anet$ is a well-structured transition system (WSTS)~\cite{FINKEL1990144} w.r.t. $\quasi$. Since $\quasi$ is decidable for the constraint language we consider and $LTS_\anet$ is a WSTS, the following holds:

    \begin{proposition}[\cite{Suvorov2024}]
\label{prop:bounded-c-graph}
    Let $\anet$ be a DPN with guards constructed from the language of arithmetic constraints as per Definition~\ref{def:constraint-syntax}.
    Let $LTS_\anet$ be an LTS as per Definition~\ref{def:LTS_new} with quasi-ordering $\quasi$.
    Then $CG_{LTS_\anet}$ is finite and effectively constructible. 
\end{proposition}
It is crucial that the above statement holds only for DPNs with the said guard language and with variables evaluated over $\Reals$. The same result does not already hold if the variables are evaluated over $\mathbb{N}$ or $\mathbb{Z}$~\cite{Suvorov2024}.

    From the above proposition, it is easy to see that the boundedness check for (the said class of) DPNs 
    is \emph{decidable} and can be effectively done by analyzing the WSTS coverability graph for the presence of strictly covering nodes~\cite{Finkel01}. 
    
\subsection{Automating the Process of Repairs}
\label{sec:process-repairs}

Let us start by introducing a computation structure suitable for automating the process of repairs. More specifically, we introduce a color-based refinement of coverability graphs from Definition~\ref{def:c-graph}.

\begin{definition}[Colored Coverability Graph]
\label{def:cc-graph}
     A \emph{colored coverability graph} (CCG) $CG_{LTS_\anet}^{c} = (S_{CG}, E_{CG}, s_0, c)$ is a coverability graph $(S_{CG}, E_{CG}, s_0)$ of a DPN $\anet$ with a final state $M_F$ that is enriched with the color function $c:S_{CG}\to\set{\mathtt{red},\mathtt{green}}$.
    For each state $s=(M,A)\in S_{CG}$, $c(s)=\mathtt{green}$ if one of the following conditions holds:
    \begin{inparaenum}[\it (i)]
    \item $M = M_F$;
    \item $s$ has a path to a green node.
    \end{inparaenum}
    Otherwise, $c(s)=\mathtt{red}$.
\end{definition}

    In the CCG, the states from which the final marking can be reached are colored in \emph{green}, and the states that lead to deadlocks, livelocks, and/or token growth are colored in \emph{red}. In the context of a soundness repair procedure, the transition firings leading from a green node to a red one are prime candidates for being prohibited.

    \begin{proposition}
    \label{prop:cc-graph-construction}
        Let $(S_{CG}, E_{CG}, s_0)$ be a coverability graph of a DPN $\anet$ with a final state $M_F$.         
        Then $CG_{LTS_\anet}^{c}$ can be effectively constructed. 
    \end{proposition}
    \begin{proof}
        To construct $CG_{LTS_\anet}^{c}$, we have to iteratively define the color function $c$.
        This can be naively done for every state in $S_{CG}$ by either checking whether its marking component coincides with $M_F$ or by running a reachability query on a finite graph in order to satisfy condition (ii) of the color function from Definition~\ref{def:cc-graph}.
    \end{proof}

    For some cases, it is enough to construct a CCG for a source DPN and forbid executions that lead to red nodes to make the net sound. In the next subsection, we present a sample unbounded DPN for which it is true.
    However, in other cases this approach does not work: the CCG structure allows detecting sources of unboundedness, but it cannot identify most DPN deadlocks and livelocks. For instance, the CCG for the DPN from Figure~\ref{fig:casino} only contains green nodes although this net is unsound (this is true since from each LTS node the final node is reachable). 
    However, we can address this problem by transforming the source net in such a way that the CCG contains refined execution paths allowing us to identify livelocks and deadlocks for the original DPN.
    For these purposes, we will first construct a \emph{refined DPN} and then convert it to \emph{tau-DPN} following algorithms defined in \cite{Suvorov2024}. We provide intuitive definitions of both such DPN types below. 
    In~\cite{Suvorov2024}, it is shown that an LTS constructed for the tau-refined DPN captures all the sources of unsoundness.

    Let $\anet$ be some DPN. A refined DPN, denoted by $\anet_{R}$, is a net that is behaviorally equivalent to $\anet$ (their reachability graphs are equivalent) and that is constructed using the algorithm presented in~\cite{Suvorov2024}. In short, the algorithm detects all the cycles occurring in $LTS_\anet$ and splits the transitions included in these cycles based on the guards of each transition leading out of these cycles.
    The described procedure is decidable for bounded DPNs in our setting.
    Specifically, if $t$ is some transition in cycle $c$, and $t_{out}$ is a transition leading out of the $c$, then $t$ is split into $t^+$ and $t^-$, where the guard of $t^+$ is a conjunction of $guard(t)$ and the input condition of $t_{out}$ and the guard of $t^-$ is a conjunction of $guard(t)$ and the negation of the input condition of $t_{out}$. The refinement is done for each transition with $write(t)\neq\emptyset$ in each cycle and is performed iteratively until the net stabilizes (i.e. none of the transitions change). The refinement is an important step that allows to capture all the livelocks of the source DPN in the LTS or CG constructed for the tau-refined DPN.

Now, we introduce yet another type of constructive modification of a DPN -- a \emph{tau-DPN}. 
In a nutshell, given a DPN $\anet$, we can obtain a tau-DPN $\anet_\tau$ out of it by enriching it with the following $\tau$-transitions: for each $t \in T$ with $guard(t)$ containing non-trivial input constraints (that is, constraints including variables from  $V^r$),
we introduce a transition $\tau_t$ with $guard(\tau_t)$ set to $\neg(\exists write(t):guard(t))$. 
Constructing an LTS or CG for $\anet_\tau$ allows detecting sets of DPN states from which the final marking is not reachable. We refer the reader to~\cite{Suvorov2024} for the detailed definition of tau-DPN.

\begin{figure}[htp]	
\centering
\resizebox{\columnwidth}{!}{
	\begin{tabular}{cc}
			\begin{tabular}{@{}c@{}}
					
     \includegraphics[width=.7\textwidth]{output-figure0.pdf}
				\end{tabular}
			&
			\begin{tabular}{@{}c@{}}
					
     \includegraphics[width=.7\textwidth]{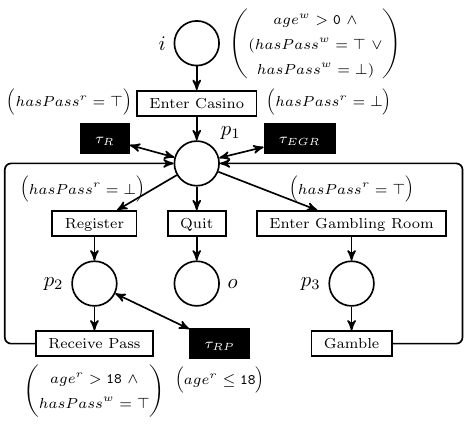}
				\end{tabular}
			\\[-3ex]
			\subfloat[]{\hspace{4cm}}
			&
			\subfloat[]{\hspace{4cm}}
		\end{tabular}
}
\caption{Constructive DPN modification needed to capture all sources of unsoundness in an LTS. (a) DPN $\anet$ from Figure~\ref{fig:casino}. (b) Modified DPN $\anet_{R_\tau}$, where $\tau_R, \tau_{EGR}, \tau_{RP}$ are $\tau$-transitions constructed for \textit{Register, Enter Gambling Room} and \textit{Receive Pass}, respectively.}
\label{fig:refined_casino}
\end{figure}

\begin{example}
    \label{ex:tau-refined}
    Figure~\ref{fig:refined_casino} illustrates the construction of the tau-refined DPN for the DPN representing a visit to a casino. Here, the refinement does not produce any new transitions since the only DPN transition that occurs in a cycle and updates a variable, \textit{Receive Pass}, conducts a deterministic transformation of a variable value (assigns $\top$ to \textit{hasPass}) and thus cannot be anyhow split based on this variable assignment. $\tau$-transitions are added for \textit{Receive Pass}, \textit{Register}, and \textit{Enter Gambling Room} as they have input conditions. The constraints of the resulting $\tau$-transitions are negations of input conditions of the source transitions.
\end{example}

Let us define the procedure that repairs DPN soundness, denoted~\textit{RepairDPN}, constructively. It takes as an input DPN $\anet$ with initial state $(M_I,\alpha_I)$ and final marking $M_F$ and returns a tuple $(\anet, isSuccess)$, where $isSuccess$ is a flag denoting whether the procedure succeeded to repair the DPN and $\anet$ is the repaired net if $isSuccess$ is true, or the source net otherwise.
The first step is to make the net bounded. For this, we construct a CCG for $\anet$ and call procedure~\textit{MakeRepairStep} defined below (Proposition~\ref{prop:repair-step-properties} shows that \textit{MakeRepairStep} always returns a bounded net).
The second step is to forbid executions that lead to deadlocks and livelocks. For this, in a loop, we construct a CCG for tau-refined DPN $\anet_{R_\tau}$ and call \textit{MakeRepairStep} if the CCG contains both green and red nodes. The exit condition for a loop is the absence of either green or red nodes. If the CCG has only green nodes, the repair is successful ($isSuccess$ becomes $\true$) and we proceed to the last step. If the CCG has only red nodes, the repair is not successful ($isSuccess$ becomes $\false$) and the algorithm terminates returning the source DPN.
The last step is executed if the repair is successful. This step removes dead transitions exploiting the information from the CCG and isolated places in $\anet$ and merges back transitions that were refined during the second step when constructing $\anet_{R_\tau}$. The modified $\anet$ is returned as a result of~\textit{RepairDPN}.

For some DPNs, the subsequent application of repair steps leads to a DPN with a colored coverability graph containing only red nodes. For these DPNs, our algorithm terminates but fails to repair soundness. The examples of such nets are presented in Section~\ref{sec:applicability}.

Procedure~\textit{MakeRepairStep} is also defined constructively. It takes as an input DPN $\anet$ and its CCG and returns a modified DPN, where some of the transition guards are restricted. Let $T_\tau$ be a set of $\tau$-transitions of $\anet$. If $\anet$ has no $\tau$-transitions, this set is empty.
The first step is to identify critical arcs in the CCG: an arc is called critical if its source node is green and its target node is red.
Critical arcs should be forbidden to make the net sound.
The second step is to identify the transitions that should be restricted and to restrict them. For each critical arc $(s,t,s')$:
\begin{compactenum}
    \item If $t\notin T_\tau$, we restrict $guard(t)$ by conjuncting it with the negation of $s'$ constraint. 
    \item If $t\in T_\tau$, we find all the nearest incoming non-tau arcs in the CCG and restrict the corresponding transitions. For this, we define $P$ as the set of all simple paths in the CCG leading to $s$. For each $p\in P$, we take the last arc $(s'',t',s''')$, such that $t'\notin T_\tau$, and add $t'$ to the set of transitions to be restricted. For each such $t'$, we restrict $guard(t')$ by conjuncting it with the negation of $s'$ constraint.
\end{compactenum}
Lastly, in each restricted guard, if $v\in write(t)$, $v$ is replaced with $v^w$, otherwise $v$ is replaced with $v^r$ to make guards the formulas of $\Phi(V^r\cup V^w)$.

\begin{figure}[!ht]
    \centering    					
    \includegraphics[width=.95\textwidth]{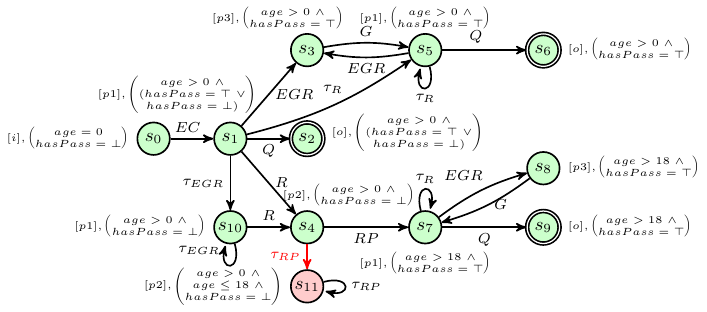}
    \caption{$CG^c$ constructed for DPN $\anet_{R_\tau}$ from Figure~\ref{fig:refined_casino}. Nodes are colored w.r.t. the color function. Red arcs denote critical arcs.}
    \label{fig:casino_dpn_refined_ccg}
    \end{figure}

\begin{figure}[!ht]	
    \centering    			
    \includegraphics[width=0.65\textwidth]{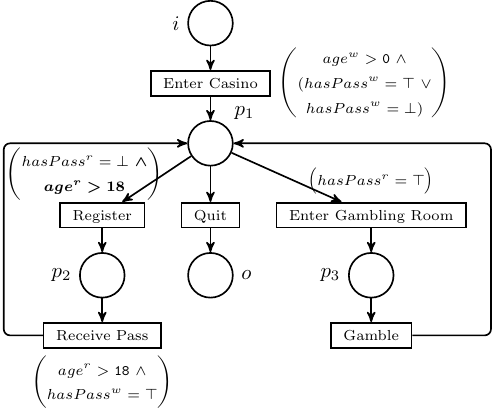}
    \caption{Repaired DPN from Figure~\ref{fig:casino}. The guard of \textit{Register} is restricted by conjuncting it with $age^r > 18$ (highlighted in bold).}
    \label{fig:casino_dpn_repaired}
\end{figure}

\begin{example}
    Figure~\ref{fig:casino_dpn_refined_ccg} demonstrates the CCG constructed for $\anet_{R_\tau}$ from Figure~\ref{fig:refined_casino}. Node $s_{11}$ represents the set of states at which the model meets a deadlock. The only critical arc in this graph is $(s_4,\tau_{RP}, s_{11})$. According to \textit{MakeRepairStep}, we need to find all simple paths that lead to $s_4$ since $\tau_{RP}$ is a tau-transition. It is easy to see that all such simple paths end with transition \textit{Register}.
    Since it is not a tau-transition, its guard should be restricted by conjuncting $guard(Register)$ with $age^r > 18~\lor~hasPass = \top~\lor~age^r \le 0$, which results in $hasPass^r = \bot \land age^r > 18$ after simplification. The CCG for the resulting net contains only green nodes, which means that the conducted restriction made the model sound. The repaired model is shown in Figure~\ref{fig:casino_dpn_repaired}. From the domain perspective, we have eliminated a potential deadlock by prohibiting registration for a pass for people not greater than 18 years old.
\end{example}

\subsection{Other examples of repair algorithm application}

    In this subsection, we showcase the application of our repair algorithm to DPNs with other sources of unsoundness. Specifically, we consider a DPN with a livelock (see Example~\ref{ex:livelock}) and a DPN with an unbounded place (see Example~\ref{ex:unbounded}).

    \begin{figure}[htp]	
\centering
\resizebox{\columnwidth}{!}{
	\begin{tabular}{cc}
			\begin{tabular}{@{}c@{}}
					
     \includegraphics[width=.4\textwidth]{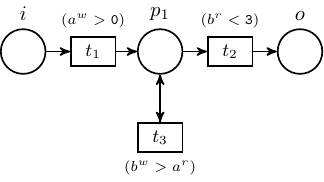}
				\end{tabular}
			&
			\begin{tabular}{@{}c@{}}
					
     \includegraphics[width=.6\textwidth]{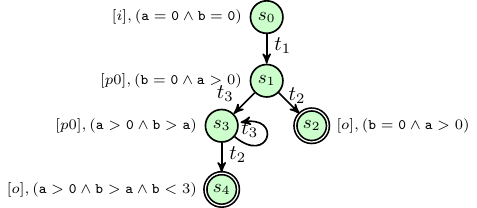}
				\end{tabular}
			\\[-3ex]
			\subfloat[]{\hspace{4cm}}
			&
			\subfloat[]{\hspace{4cm}}
		\end{tabular}
}
\caption{
Livelock example.
(a) DPN $\anet$ with a livelock at $M=[p_1]$, $a \ge 3$ and $b > 3$. $M_I=[i]$ and $M_F=[o]$. $\varInit(a)=\cval{0}$ and $\varInit(b)=\cval{0}$. 
(b) $CG^c$ constructed for DPN $\anet$. Nodes are colored w.r.t. the color function.}
\label{fig:fail_dpn}
\end{figure}

\begin{figure}[htp]	
\centering
\resizebox{\columnwidth}{!}{
	\begin{tabular}{cc}
			\begin{tabular}{@{}c@{}}
					
    \includegraphics[width=.35\textwidth]{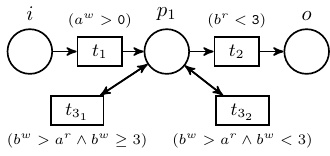}
				\end{tabular}
			&
			\begin{tabular}{@{}c@{}}
					
     \includegraphics[width=.35\textwidth]{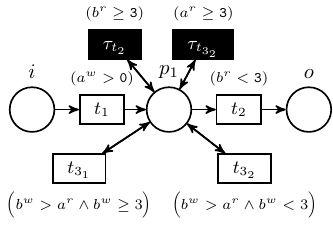}
				\end{tabular}
			\\[-3ex]
			\subfloat[]{\hspace{4cm}}
			&
			\subfloat[]{\hspace{4cm}}
		\end{tabular}
}
\caption{
Livelock example: transformations.
(a) Shows a refined DPN $\anet_{R}$, where $t_{3_1}$ and $t_{3_2}$ are transitions resulted from splitting $t_3$. 
(b) Shows a tau-DPN $\anet_{R_\tau}$, where $\tau_{t_2}$ and $\tau_{t_{3_2}}$ are $\tau$-transitions for $t_2$ and $t_{3_2}$, respectively.}
\label{fig:refined_dpn}
\end{figure}

\begin{figure}[htp]	
\centering
\resizebox{\columnwidth}{!}{
	\begin{tabular}{cc}
			\begin{tabular}{@{}c@{}}
					
     \includegraphics[width=.6\textwidth]{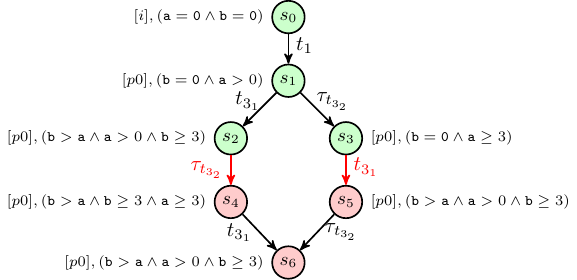}
				\end{tabular}
			&
			\begin{tabular}{@{}c@{}}
					
     \includegraphics[width=.4\textwidth]{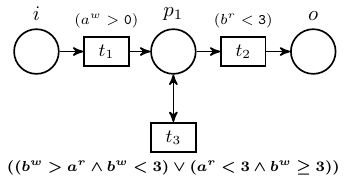}
				\end{tabular}
			\\[-3ex]
			\subfloat[]{\hspace{4cm}}
			&
			\subfloat[]{\hspace{4cm}}
		\end{tabular}
}
\caption{
Livelock example: repair result.
(a) A fragment of $CG^c$ for $\anet_{R_\tau}$ from Figure~\ref{fig:refined_dpn}(b) containing only paths from the initial state to a red node. Red arcs denote critical arcs.
(b) Repaired DPN $\anet$ from Figure~\ref{fig:fail_dpn}. Changes are highlighted in bold.}
\label{fig:fail_result_dpn}
\end{figure}

\begin{example}
    Figure~\ref{fig:fail_dpn} shows DPN $\anet$ having a livelock at $M=[p_1]$ when $a \ge 3$ and $b > 3$ and its CCG. Since the CCG does not have red nodes, the first step of \textit{RepairDPN} does not anyhow change the DPN. At the second step, the tau-refined DPN is constructed which is shown in Figure~\ref{fig:refined_dpn}. Transition $t_3$ is split into $t_{3_1}$ and $t_{3_2}$ based on the input condition of transition $t_2$. $\tau$-transitions are only added for $t_2$ and $t_{3_2}$ as other transitions do not have input conditions. A fragment of the CCG for $\anet_{R_\tau}$ is illustrated in Figure~\ref{fig:fail_result_dpn}(a). Here, the critical arcs are $(s_2,\tau_{t_{3_2}},s_4)$ and $(s_3,t_{3_1},s_5)$. According to the logic of~\textit{MakeRepairStep}, the only transition that should be restricted is $t_{3_1}$. We need to add negations of the constraints of $s_4$ and $s_5$ to its guard so that the new guard (after simplifications) becomes $(b^w \ge 3 \land a^r < 3)$. This concludes the first iteration of the loop. 
    The CCG constructed on the second iteration has only green nodes; thus, we proceed to the next step, on which transitions $t_{3_1}$ and $t_{3_2}$ are merged into $t_3$, whose guard becomes a disjunction of guards of $t_{3_1}$ and $t_{3_2}$. Since the DPN does not contain dead transitions and isolated places, no other changes to the DPN are made. The repaired DPN is presented in Figure~\ref{fig:fail_result_dpn}(b).
\label{ex:livelock}
\end{example}

\begin{figure}[!ht]	
\centering
\resizebox{\columnwidth}{!}{
	\begin{tabular}{cc}
			\begin{tabular}{@{}c@{}}
					
     \includegraphics[width=.55\textwidth]{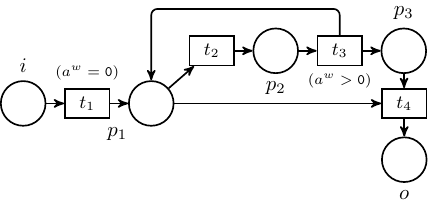}
				\end{tabular}
			&
			\begin{tabular}{@{}c@{}}
					
     \includegraphics[width=.4\textwidth]{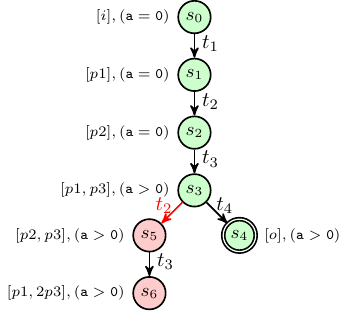}
				\end{tabular}
			\\[-3ex]
			\subfloat[]{\hspace{4cm}}
			&
			\subfloat[]{\hspace{4cm}}
		\end{tabular}
}
\caption{
Unbounded example.
(a) DPN $\anet$ with unbounded place $p3$. $M_I=[i]$ and $M_F=[o]$. $\varInit(a)=\cval{0}$. 
(b) $CG^c$ constructed for DPN $\anet$. Nodes are colored w.r.t. the color function. Red arcs denote critical arcs.}
\label{fig:unbounded_dpn}
\end{figure}

\begin{figure}[!ht]	
    \centering    			
    \includegraphics[width=.5\textwidth]{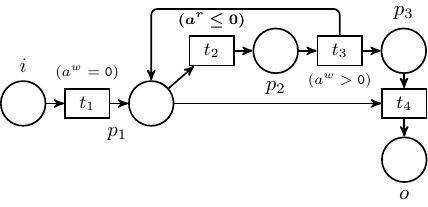}
    \caption{Unbounded example. Repaired DPN $\anet$ from Figure~\ref{fig:unbounded_dpn}(a). Changes are highlighted in bold.}
    \label{fig:unbounded_dpn_result}
\end{figure}
    
    \begin{example}
        Figure~\ref{fig:unbounded_dpn} shows a DPN $\anet$ with an unbounded place $p_1$ and its CCG. Since the net is unbounded, the CCG must include red nodes. The only critical arc here is $(s_3,t_2,s_5)$. Therefore, the guard of $t_2$ is modified by conjuncting $guard(t_2)$ and negation of the constraint of $s_5$. The guard of $t_2$ becomes $a \le 0$. This modification makes the DPN bounded. The loop of procedure~\textit{RepairDPN} is executed only once as $\anet$ does not have deadlocks or livelocks. The repaired DPN is presented in Figure~\ref{fig:unbounded_dpn_result}.
        \label{ex:unbounded}
    \end{example}

    \subsection{Main Algorithm Properties}

Below, we formulate the most important properties of procedure~\textit{MakeRepairStep}.

\begin{proposition}
\label{prop:repair-step-properties}
Given a \net $\anet$ with initial state $(M_I,\alpha_I)$ and final marking $M_F$,
and a colored coverability graph $CG^{c}_{LTS_\anet}$,
procedure \textit{MakeRepairStep} 
\begin{inparaenum}[\it (1)]
    \item terminates, and
    \item returns $\anet'$ such that $RG_{\anet'}$ is a sub-graph of $RG_{\anet}$ and $\anet'$ is bounded.
\end{inparaenum}
\end{proposition}
\begin{proof}
    \textit{MakeRepairStep} always terminates as it has to explore finitely many arcs, and for each of such arc it performs finitely many guard enhancements.

    The second property results from the idea that the algorithm only eliminates unwanted behaviors by enhancing the transition guards of $\anet$. We elaborate more on it below.

    Assume that the input net is unbounded. According to the algorithm, we consider only transitions from the CCG that lead from a $\mathtt{green}$ state (describes a node from which $M_F$ can be reached without accumulating tokens, i.e., they are not strictly covered) to a $\mathtt{red}$ one (describes a node leading to unboundedness, livelocks or deadlocks, and from which there is no path to $M_F$ which does not infinitely accumulate tokens in at least one place).
    Let $(s,t,s')$ with $s = (M,A)$ and $s' = (M',A')$ be some arc in $CriticalArcs$, which means that $s$ is a green node and $s'$ is a red node. If $t$ is not a $\tau$-transition, we restrict its guard; otherwise, we restrict guards of all the closest non-$\tau$-transitions. After this restriction, the updated CCG will not contain neither arc $(s,t,s')$ nor any arc $(s'',t,s''')$, where $s'' = (M, A'')$ with $A''\subseteq A$ and $s''' = (M, A''')$ with $A''' \subseteq A'$.
    After iterating over all the elements from $CriticalArcs$, the colored coverability resulting graph will not contain any node $s = (M,A)$ such that there existed $\mathtt{red}$ node $s' = (M,A')$ with $A\subseteq A'$ in the source colored coverability graph. Thus, the resulting graph will have no strictly covering states. This, in turn, means that there are no paths on this graph leading to unboundedness. 

    It is also easy to see that $RG_{\anet'}$ is a sub-graph of $RG_{\anet}$. By restricting the guards of $\anet$, we only forbid its certain execution paths that would lead to states described by $s'$. This means that new behaviors do not emerge, and the net inherits only the behaviors manifesting between the green nodes of $CG^{c}$.
\end{proof}

    We now show that procedure~\textit{RepairDPN} always terminates and that, whenever it succeeds in repairing a model, it does not introduce any new behavior. 

    \begin{proposition}
    \label{prop:termination}
        For any DPN $\anet =  \tup{P,T,F,V,\Phi^\Reals,guard}$ with initial state $(M_I, \alpha_I)$ and final marking $M_F$, procedure  
        $\textit{RepairDPN}(\anet, (M_I,\alpha_I), M_F)$ terminates.
    \end{proposition}
    \begin{proof}

        We know that $CG_{LTS_\anet}^c$ for a DPN $\anet$ is finite and can be effectively constructed (see Propositions~\ref{prop:bounded-c-graph} and \ref{prop:cc-graph-construction}) and that \textit{MakeRepairStep} always terminates. Thus, the first step of~\textit{RepairDPN} terminates.

        Let us now consider the second step of procedure~\textit{RepairDPN}.
        According to Proposition~\ref{prop:repair-step-properties}, the loop starts with a new DPN that is already bounded. 
        Each loop iteration only restricts the net's behavior by calling \textit{MakeRepairStep} if the net's CCG
        contains at least one red and one green node. 
        Each iteration of this loop terminates if the current DPN $\anet$ is bounded.
        The loop only restricts the net behavior (see Proposition~\ref{prop:repair-step-properties}), which preserves the DPN boundedness, and eventually terminates. The latter partially follows from the termination of each of its subroutines:
        the construction of $\anet_{R_\tau}$ terminates if $\anet$ is bounded~\cite{Suvorov2024} and \textit{MakeRepairStep} terminates according to Proposition~\ref{prop:repair-step-properties}. 
        The number of the loop iterations is always finite, since each transition guard may be restricted finitely many times as only a finite set of non-equivalent constraints can be constructed for each guard using the elements from $\Phi(V)$. 
        
        Finally, since the set of places and transitions is finite for any DPN, the procedures of removing dead transitions, removing isolated and merging refined transitions terminate. 
        Thus, \textit{RepairDPN} terminates.
    \end{proof}
    
    Next, we show that a repaired DPN does not allow any new behavior that was not present in the source net:

    \begin{proposition}
    \label{prop:restriction}
        Let $\anet =  \tup{P,T,F,V,\Phi^\Reals,guard}$ be a DPN. Let $\anet_{Rep}$ be a repaired DPN, obtained by executing \textit{RepairDPN} on $\anet$. Then $RG_{\anet_{Rep}}$ is a subgraph of $RG_{\anet}$.
    \end{proposition}
    \begin{proof}
        Procedure \textit{RepairDPN}, at each iteration of the loop, splits DPN transitions and restricts guards of the resulting transitions. 
        After the loop, all the split transitions are merged back.
        Note that all these operations preserve pre- and post-sets of transitions; thus, it is only meaningful to estimate their influence on the DPN behavior in terms of changes in transition guards.
        Let $t$ be some DPN transition. Let $t_1,...,t_n$ be transitions resulted from splitting $t$. 
        It is easy to see that merging transitions and/or splitting them back does not add or remove any new behavior.
        Consequently, only restrictions imposed on transition guards affect the DPN's behavior. 
        The algorithm restricts $guard(t)$ by substituting $guard(t)$ with $guard_{res}(t)$, where $guard_{res}(t)$ is a conjunction of $guard(t)$ and some arithmetic constraints $c_1,...,c_m$. Consequently, $[[guard_{res}(t)]]\subseteq [[guard(t)]]$, which means that restricting a transition guard does not add any new behavior (due to the subsumption of sets of all possible variable assignments satisfying the respective guards). 
        As a result, splitting transitions, restricting transition guards, and merging transitions do not add any new behavior. 
        Since removing dead transitions and isolated places, which is done at the end of procedure~\textit{RepairDPN}, also cannot add any new behavior to the net, $RG_{\anet_{Rep}}$ is always a subgraph of $RG_{\anet}$.
    \end{proof}

Notice that \textit{RepairDPN} is a decision procedure: given a DPN, it determines whether the net can be repaired.
As previously discussed, \textit{RepairDPN} always terminates. If it provides a positive result, it also outputs a repaired DPN that is guaranteed to be sound (soundness for the type of DPNs considered in this paper can always be verified using the procedure outlined in~\cite{Suvorov2024}).
However, if the algorithm is unable to produce a repaired model, it returns the original model along with a negative result. It is important to note that this negative result can be a ``false negative'', as \textit{RepairDPN} is not guaranteed to repair every input net.
This makes \textit{RepairDPN} a \emph{semi-decision procedure}.

\section{Implementation and Experiments}
\label{sec:implementation}

\begin{figure}[!ht]
\centering
\includegraphics[width=.9\textwidth]{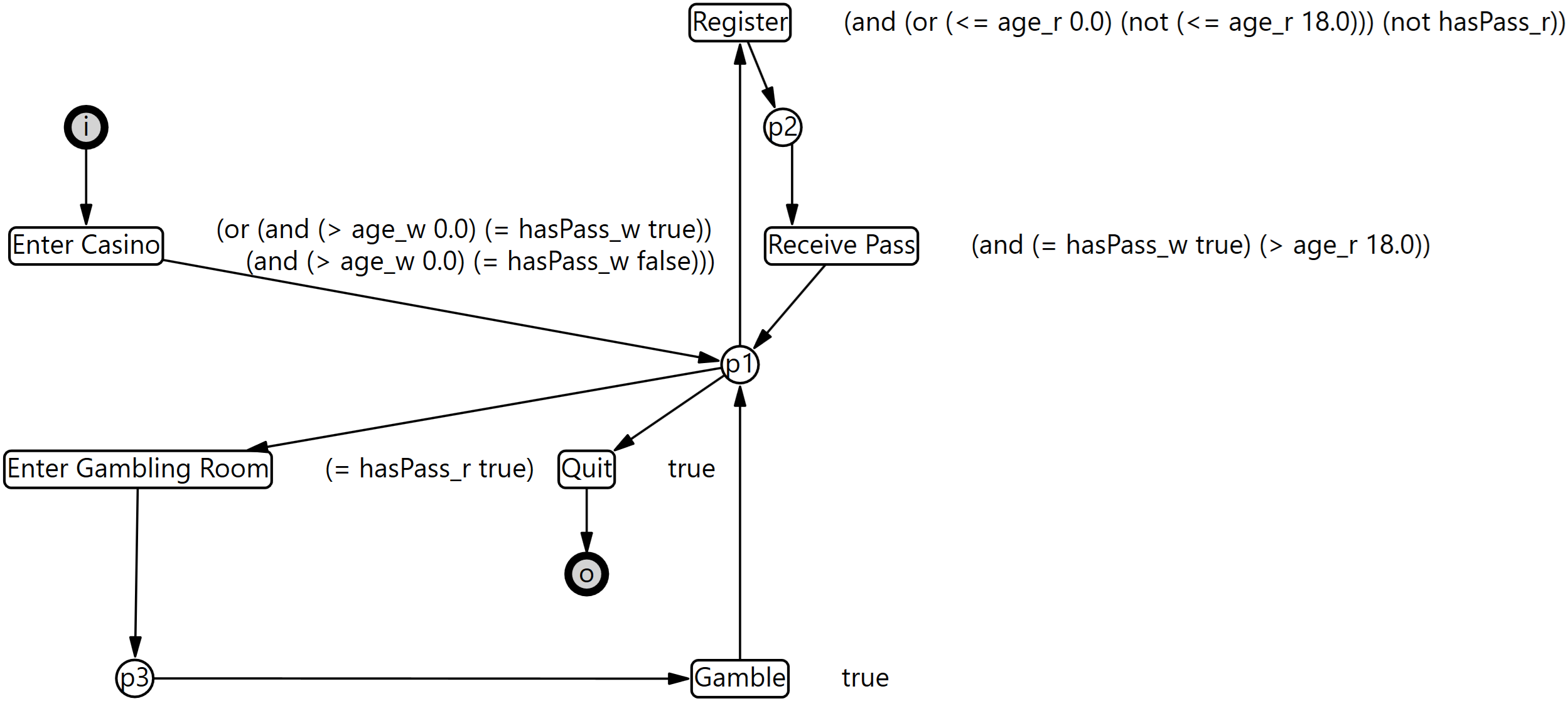}
\caption{A result of repairing DPN $\anet$ from Figure~\ref{fig:casino} using the implemented algorithm.}
\label{fig:repair_result}
\end{figure}

The proposed algorithm for data-aware soundness repair 
has been implemented as a module in the existing DPN soundness verification tool implemented on .NET WPF. 
The application with the repair module is available for download on Github\footnote{https://github.com/SuvorovNM/DPN-Soundness-Verification}. 
As an example, Figure~\ref{fig:repair_result} shows how the implemented toolkit repaired the DPN from Figure~\ref{fig:casino}. 
The resulting DPN is equivalent to the manually repaired DPN presented in Figure~\ref{fig:casino_dpn_repaired} except for the fact that the implemented tool has not simplified the guard of~\textit{Register}. Nonetheless, the reachability graphs for both of these nets are equivalent.

At the implementation level, the following small adjustment to the algorithm has been done to decrease the repair time. 
The refinement is performed only if it is the first iteration of the loop or if at the previous iteration a CCG with all green nodes has been obtained (the exit condition for the loop is changed to having a CCG with only green nodes at the previous iteration and a CCG with only green nodes at the current iteration). 
The DPN refinement is a time-consuming procedure; thus, it is reasonable to postpone the refinement if it is possible. This helps to significantly decrease the time needed for the repair.
At the same time, implementation-specific changes made to \textit{RepairDPN} do not affect the properties studied for the algorithm in Section~\ref{sec:repair_algorithm}.

We have evaluated the performance of the developed algorithms on synthetic and real-life data. All the experiments have been conducted on Intel Core i7-12700H with 16 GB RAM. 
Repair of each DPN has been conducted three times to get a real average neglecting the impact of the external factors on the repair time.

First, we have decided to compare our algorithm with the algorithm from~\cite{Felli_Repair} that follows the same approach that we investigate in this paper, i.e. restricts the transition guards. However, it is important to note that the algorithms have slightly different scopes of use. Algorithm~\cite{Felli_Repair} is proposed to be mainly used for the models obtained during the process discovery (thus, control flow soundness is presumed) whose transition guards can be composed of arithmetic conditions. Our algorithm is proposed to be mainly used for the models constructed manually (thus, control flow soundness is not presumed) and the transition guards of the models can only be composed of variable-operator-constant and variable-operator-variable conditions.

Table~\ref{tab:repairTime} reports on how much time and how many repair steps the compared algorithms take to repair soundness of different DPN models. 
Generally, our algorithm allows to repair soundness of a DPN quicker than the algorithm from~\cite{Felli_Repair} although the number of repair steps can be higher. 
This may have different underlying reasons. 
First, the underlying verification algorithms are different, whereas in repair algorithms, most of the repair time is wasted on estimating DPN soundness.
In~\cite{Suvorov2024}, we showed that soundness verification based on constructing an LTS for a tau-refined DPN (the approach used in our algorithm) usually has slightly better time results than the verification based on constructing a constraint graph for each reachable DPN marking (the approach used in~\cite{Felli_Repair}). 
Second, the languages used to implement the tools are different: C\# is used for our algorithm, and Python is used for algorithm~\cite{Felli_Repair}. With the same algorithm, implementation in C\# will often be faster.
Third, the algorithm~\cite{Felli_Repair} is proposed for the more general case in terms of available transition guards and this could also impact both the verification and the repair time.

\begin{table}[htbp]
\caption{Soundness repair time for sample DPNs}
\centering
\resizebox{.85\textwidth}{!}
{\begin{tabular}{| m{5.1cm} | m{2cm}|  m{2cm} |m{2cm}|m{2cm}|}
\hline
\centering Model & \textit{RepairDPN} Repair Time & \textit{RepairDPN} Repair Steps & Algorithm~\cite{Felli_Repair} Repair Time & Algorithm~\cite{Felli_Repair} Repair Steps \\
\hline
Casino Example (Figure~\ref{fig:casino}) & 169 ms & 2 & 2.7 s & 2 \\
\hline
Livelock Example (Figure~\ref{fig:fail_dpn}) & 203 ms & 1 & 2.1 s & 1 \\
\hline
Unbounded Example (Figure~\ref{fig:unbounded_dpn}) & 40 ms & 1 & - & - \\
\hline
Digital Whiteboard: Transfer~\cite{Mannhardt_phd} & 143 ms & 4 & 2.1 s & 1 \\
\hline
Package Handling~\cite{Felli2021} & 4.8 s & 0 & 6 s & 0 \\
\hline
Road Fines Mined~\cite{Mannhardt_phd} & 1.9 s & 1 & 24 s & 1 \\
\hline
Simple Auction~\cite{Felli_Repair} & 318 ms & 1 & 2.5 s & 1 \\
\hline
\end{tabular}}
\label{tab:repairTime}
\end{table}

\begin{figure}[!htp]
\centering
\includegraphics[width=.9\textwidth]{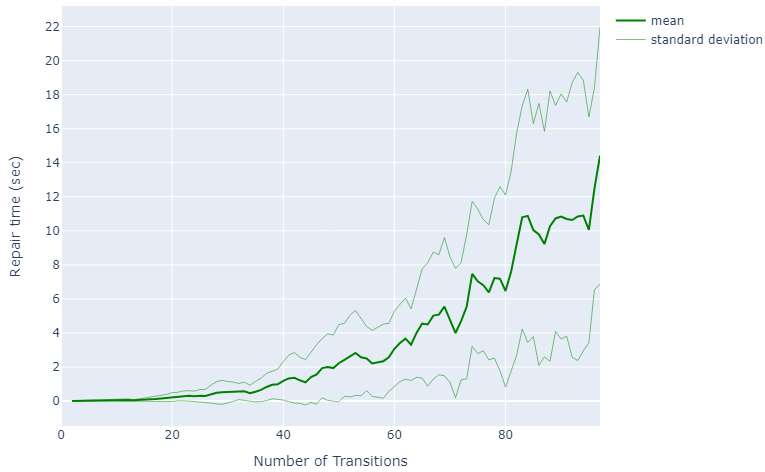}
\caption{\textit{RepairDPN} procedure execution time on bounded DPNs of different sizes}
\label{fig:repair_time}
\end{figure}

We have also conducted other experiments to evaluate the performance of the algorithm.
Given $n \in \nat$, we considered DPNs parameterized according to the following setup:
\begin{compactitem}
\item $1.2n$ places,
\item $n$ transitions,
\item $0.25n$ variables, and
\item $0.5n$ conditions.
\end{compactitem}
For each $n\in\nat$ from 3 to 100, we generated 10 DPNs that have at least one trace leading to $M_F$ using the tool introduced in~\cite{Suvorov2024}. This tool conducts three steps to generate a DPN. First, it generates a net with a sound control flow based on the defined numbers of places and transitions. Second, it adds extra arcs to the net. Third, it generates random formulas according to the numbers of variables and atomic conditions and puts them on DPN transitions.
On each DPN, we executed our repair algorithm. The obtained results are visualized in Figure~\ref{fig:repair_time}. The plot shows that our algorithm generally requires less than half a minute to repair a DPN with less than 100 transitions. If the Road Fines model presented in~\cite{Mannhardt_phd} is considered as a small model, then we can say that our algorithm is applicable for process models of both small and medium sizes. 

Note that in the worst-case scenarios, the repair time can be much higher. 
According to the complexity analysis of the verification algorithm conducted in~\cite{Suvorov2024}, we suppose that the worst-case models for our repair algorithm given the fixed number of places and transitions are those that have the largest formulas on transitions and that have as many cycles as possible. In such models, the DPN refinement can lead to a substantial increase in the length of formulas and, thus, in the time of operations on formulas.
In the future, we plan to develop the tools for generating such models to evaluate our algorithm on them.

Nevertheless, given the preliminary experimental evaluation, we believe that the obtained results for the DPN repair algorithm are promising as, for most of the cases,
nets used in practice are of small or medium sizes and have a fairly small number of formulas and cycles and, thus, the algorithm would be able to terminate in less than 30 seconds.

\section{Related Work}
    \label{sec:related_work}

   It is often the case when the manually created process models have errors. 
   Different papers, such as~\cite{ErrorOccurrence, EmpiricalStudy1, EmpiricalAnalysis2}, analyze the sources of such errors and the reasons why they are made. The research~\cite{MENDLING2008312} has investigated the SAP reference model expressed in~\cite{Curran1997SAPRB,Keller1998SAP} and found that at least 5.6\% of the included process models contain errors. The study~\cite{EmpiricalAnalysis2} has explored industrial process models and found that more than 72\% of the models used in practice have errors. Although model errors may be of different types, some errors could be detected during the soundness verification procedure. However, the found errors should somehow be fixed, which is usually a far more challenging task. A process of fixing errors that are found during the soundness verification is called soundness repair.
   Soundness repair is not significantly investigated, partly due to the complexity of this task. Most of conducted research is done specifically for classical Petri nets but the existing works still have significant limitations. 
    
    Lomazova et al. have investigated live and unbounded Petri nets and proposed algorithms to control the behavior of a process making this process bounded~\cite{Lomazova_Priorities, Lomazova_Time}. In~\cite{Lomazova_Priorities}, they explore cycles that contain all transitions in a net, construct a spine-based coverability tree based on the detected cycles and compute a priority relation on transitions that allows to forbid runs leading to unboundedness. The algorithm returns a Priority Petri net as a result. In~\cite{Lomazova_Time}, a similar approach is used but instead of priorities, time constraints are applied and, therefore, a Time Petri net is returned as a result. For both of these algorithms, termination is not guaranteed. 
    
    Gambini et al.\cite{Gambini_ErrorCorrection} proposed a heuristic optimization algorithm for repairing a net.
    At each iteration, the algorithm performs different types of small changes, compares the costs of these actions,
    and chooses the most promising result for further steps. The result of the algorithm is a set of repaired Petri nets constructed based on the source one. However, the algorithm does not guarantee that the resultant nets are always sound and, thus, they may contain errors. Another limitation of this algorithm is its underlying structure which requires a significant amount of time and space to conduct the repair.
    
    Ramezani et al.~\cite{Sidorova_Repair} considered workflows extended with resources and proposed an algorithm to repair a workflow that is unsound from the resource perspective by synthesizing a controller so that the composition of the workflow and the controller becomes sound. The synthesized controller regulates transitions that produce or consume tokens from a resource place, thereby managing the order in which certain tasks may occur and preventing the workflow from getting stuck. However, the algorithm assumes that the net is bounded and, thus, it only eliminates deadlocks and livelocks occurring in the net.
    
    Awad et al.~\cite{Awad_Repair} focused on the interplay between control and data flows and introduced an algorithm that detects and repairs data anomalies in Petri nets according to the predefined strategies. 
    An input model for this algorithm is a Petri net, for which the final marking is reachable from each reachable marking; thus, the algorithm cannot be used to repair control flow anomalies.
    This algorithm can only be used for a small subset of data-aware processes as processes with conditions over infinite domains, such as integer and real numbers, cannot be modeled with classical Petri nets.

    Regarding the soundness repair of DPNs, only two works dedicated to this topic have been found in the literature at the time of writing. In the following paragraphs, we describe in detail the approaches that they propose.

    The algorithm proposed by Zavatteri et al.~\cite{Zavatteri_2024} allows to repair soundness of an acyclic DPN that has a sound control flow and atomic conditions on transitions. Each atomic condition has the form $(x - y \circ Z)$, where $x,y$ are variables and $Z$ is a constant. The algorithm is based on the idea of small perturbations proposed in~\cite{Gambini_ErrorCorrection}. The algorithm only changes the transition guards and selects the transition guards that should be updated exploiting information from the constraint graph. As a cost function, the authors use the number of transition guards that differ from the guards of the source net. 
    Nevertheless, the algorithm has a narrow scope of use due to the constraints on DPN acyclicity and control flow soundness.

    Felli et al.~\cite{Felli_Repair} proposed an algorithm that repairs soundness of DPNs with a sound control flow and arithmetic conditions on transitions. The authors base their approach on restricting or relaxing transition guards to make the final model sound. For these purposes, they construct and analyze different types of constraint graphs of a DPN and take corresponding actions either to forbid unfeasible runs or to continue them to the proper ending. The authors assume that the unsoundness is caused only by adding the data perspective to a Petri net. Thus, the net from Figure~\ref{fig:unbounded_dpn} cannot be repaired using this algorithm. Although the authors state that their restriction algorithm always succeeds in repairing soundness and preserves all the correct behavior of the source model, this is not true (a limitation of the overall restricting transition guards approach), the counterexamples can be found in Subsection~\ref{sec:applicability}. On the contrary, their relaxing algorithm indeed always succeeds. These algorithms have an open-source software implementation and the results of the experiments show that they are mainly applicable for models of small and moderate sizes. Even for small models (such as the DPN from Figure~\ref{fig:fail_dpn}), they take more than 2 seconds to conduct the repair. We guess that the reason for this is a requirement to construct an abstract state space structure for each DPN marking at each repair step, which results in the high time complexity of the algorithms.

    Among the described algorithms, only algorithms proposed in~\cite{Awad_Repair, Zavatteri_2024,Felli_Repair} can be used to repair unsound data-aware process models. However, all of them assume that the control flow of the model is sound and, thus, they focus on repairing the data flow component only. We find this assumption rather strong since even sound data-aware process models may have an unsound control flow. In addition, repair of unsound models with sound control flow can always be done by simply removing the data flow. In this work, we have tried to overcome this limitation and proposed an algorithm that is also applicable to data-aware process models with an unsound control flow.

    Our algorithm takes as an input an arbitrary DPN with real type variables and logical expressions composed of variable-operator-constant and variable-operator-variable conditions as transition guards. The proposed algorithm is designed for scenarios when a modeler properly defines the correct executions but may miss some deadlocks, livelocks, and/or unbounded resources. In these cases, no new behavior should be added to the model. The approach we have chosen to repair a model is to restrict the transition guards. By that, the executions that previously led to improper termination become forbidden. However, as we have shown in the previous sections, the approach that we follow still has significant limitations: for some models, the algorithm may either fail to repair a model or also restrict the behavior that was correct in the source model. Nevertheless, at the current state, there exists no algorithm that guarantees the success of a repair for an arbitrary Petri net; thus, the obtained results are quite expected.

\section{Conclusion}
\label{sec:conclusion}

In this paper, we have proposed an algorithm that allows to repair soundness of data-aware process models, represented by Data Petri nets, which prohibits executions leading the source model to improper termination by restricting transition guards. 
As an input, the algorithm takes a DPN that has at least one execution that leads to the final marking.

We have proved that the algorithm terminates for any DPN with real-typed variables and that it does not add any new behavior to the input model. Moreover, we have shown that the reachability graph of the repaired net is a subgraph of the reachability graph of the input net.
Although the algorithm may not succeed in repairing soundness of some DPNs due to the limitation of the chosen approach, our investigation shows that for DPNs, where the control flow is sound, the algorithm is inapplicable only if the net allows for concurrent executions on multiple threads, where at least two threads update the same variable and at least one of these threads further checks its value. We also discuss that for DPNs, where the control flow is unsound, one may obtain more cases in which it is impossible to repair a model only by restricting transition guards. A trivial example is an unsound net without any transition constraints. 

The algorithm has been implemented as a module of an existing DPN soundness verification tool. The conducted experiments have shown the practical applicability of the algorithm for repairing process models of small and medium sizes. 
We tested our algorithm on some unsound models from the literature and the examples presented in this paper. The algorithm succeeded to repair each of them in less than 5 seconds. 
The proposed algorithm can be used right after discovering or manually constructing a data-aware process model in order to repair errors occurring both at control and data levels. The algorithm can potentially be incorporated in some dialog repair systems to allow a domain expert to define whether some constraint restriction is relevant or not. In this case, our algorithm can be combined with the algorithm~\cite{Felli_Repair} that relaxes the constraints to make the repair more precise and flexible.

In the future, we plan to investigate other possibilities to repair a DPN that could guarantee a successful result on any bounded DPN without adding any new behavior to the model. 
For instance, we could allow introducing new 'silent' variables when restricting transition guards and by that achieve more control over the transition firings.
We also plan to investigate for which well-defined variants of DPNs (in which the control-flow components correspond to more restrictive Petri net sub-classes) the repair procedure can be fully decidable.

\section*{Acknowledgements}

This study has been supported by the Basic Research Program at HSE University, Russia.

\bibliography{main}

\end{document}